\documentclass[tighten]{aastex6}

\usepackage{natbib}
\usepackage{latexsym}
\usepackage{graphicx}
\usepackage{amsmath}
\usepackage{amssymb}
\usepackage{url}
\usepackage{color}
\slugcomment{White Paper for the Next Generation Solar Physics Mission}

\shorttitle{Coronal Dynamics}
\shortauthors{Morton et al.}

\begin{document}
\title{Exploring Coronal Dynamics: A Next Generation Solar Physics Mission white paper}

\author{R.~J.~Morton\altaffilmark{1}$^*$, E.~Scullion\altaffilmark{1}, D.~S.~Bloomfield\altaffilmark{1}, 
J.~A.~McLaughlin\altaffilmark{1}, S.~Regnier\altaffilmark{1}, S.~W.~McIntosh\altaffilmark{2}, S~Tomczyk\altaffilmark{2},
P. Young\altaffilmark{1,3,4}}

\affil{$^1$Department of Mathematics, Physics and Electrical Engineering, Northumbria University, Ellison Building,
Newcastle upon Tyne, NE1 8ST, UK;\\
$^2$High Altitude Observatory, NCAR, P.O. Box 3000, Boulder, CO 80307-3000, USA;\\
$^3$College of Science, George Mason University, 4400 University Drive, Fairfax, VA 22030, USA;\\
$^4$Code 671, NASA Goddard Space Flight Center, Greenbelt, MD 20771, USA\\
* Correspondence should be sent to richard.morton@northumbria.ac.uk}

\begin{abstract}
Determining the mechanisms responsible for the heating of the coronal plasma and maintaining and accelerating the 
solar wind
are long standing goals in solar physics. There is a clear need to constrain the energy, mass and momentum flux through 
the solar corona and advance our knowledge of the physical process contributing to these fluxes.  Furthermore, the 
accurate forecasting of Space Weather conditions at the near-Earth environment and, more generally, 
the plasma conditions of the solar wind throughout the heliosphere, require detailed knowledge of these fluxes in
the near-Sun corona. Here we present a short case for a space-based imaging-spectrometer coronagraph,
which will have the ability to provide synoptic information on the coronal environment and 
provide strict constraints on the mass, energy, 
and momentum flux through the corona. The instrument would ideally achieve cadences of $\sim10$~s, spatial resolution of
1" and observe the corona out to 2~$R_{\sun}$. Such an instrument will enable significant progress in our understanding of
MHD waves throughout complex plasmas, as well as potentially providing routine data products to aid
Space Weather forecasting.

%{\color{blue} White papers should include descriptions of the science objective and related observable(s), the perceived importance of the objective to the field of solar physics, the type of instrumentation that is likely to be needed, and why the observation must be done from space.
%White papers should summarise the anticipated science objectives and convey a sense of their importance, urgency, and timeliness. They should explain why those objectives cannot be accomplished under the present time allocation system and should include a preliminary assessment of the feasibility of the proposed changes.}

\end{abstract}

%\keywords{Sun, Telescopes}

\section{Introduction}
The goals of understanding the physical mechanisms behind coronal plasma heating and solar wind acceleration are still
pertinent. This is due to the potential for different mechanisms, e.g., wave dissipation, turbulence, magnetic 
reconnection, instabilities, to contribute by varying amounts to the energy flux of geometrically distinct magnetic regions 
(i.e., active regions, closed quiescent loops, open field lines). Recent international interest in the forecasting of
Space Weather has 
added a fresh impetus for making progress in the problems of heating and wind acceleration. In particular, the ability to 
makes predictions of both slow and fast solar wind stream properties, and understanding their variability, are key aspects 
for determining particle fluxes into the near-Earth environment (and heliosphere) and a contributor to the evolving 
kinematics of coronal mass ejections through the heliosphere.

\begin{figure*}
	\centering
	\includegraphics[scale=0.47, clip=true, viewport=1.0cm 5.0cm 19.cm 18.cm]{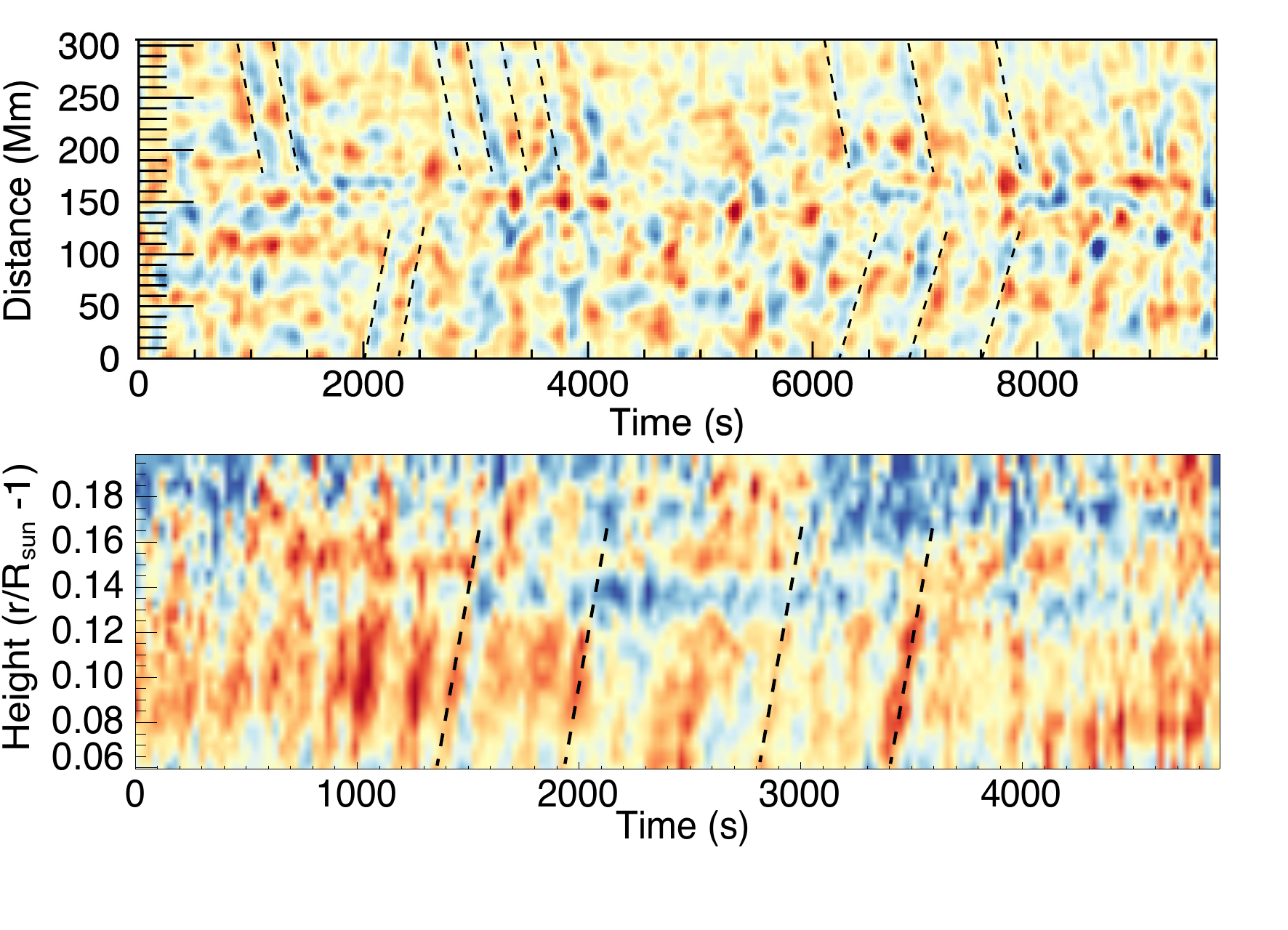}
	\includegraphics[scale=0.37, clip=true, viewport=0.0cm 6.0cm 21.cm 23.cm]{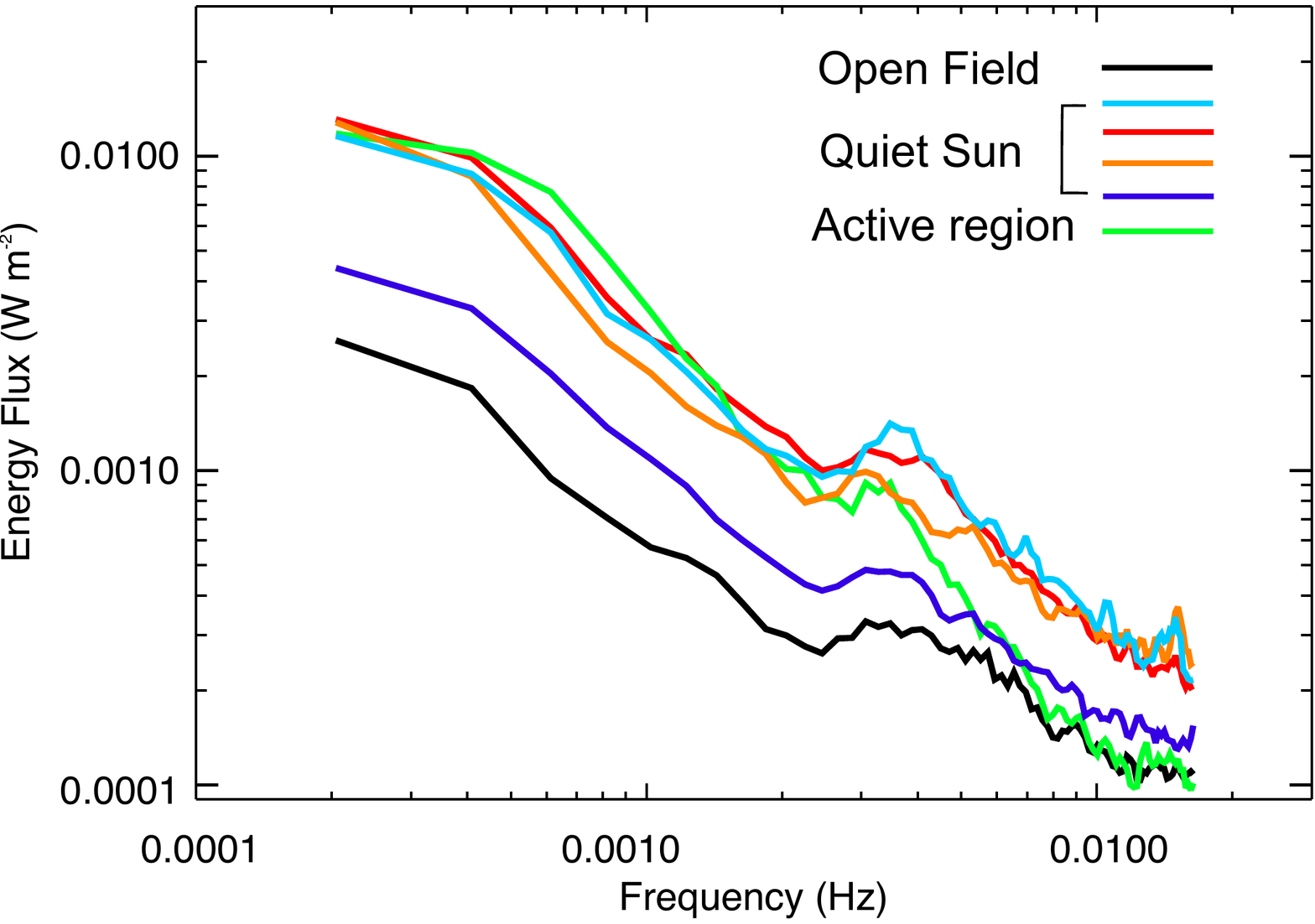}
	\caption{Measurements of the Fe XIII (1074.7~nm) line reveal a wealth of propagating, quasi-periodic 
	features in Doppler velocities throughout the solar corona, which follow the magnetic field and are interpreted 
	in terms of Alfv\'{e}nic waves. A basic result reveals somewhat balanced counter-propagating
	Doppler velocity signals in coronal loops (top left panel) and a more uneven, predominately outward 
	propagating signals along open field regions (bottom left panel). The right hand panel shows an estimate for the
	relative rates of Alfv\'enic wave energy flux for different magnetic regions in the corona (modified from
	\citealp{MORetal2016}). }
	\label{fig:features}
\end{figure*}

Recently, a potential basal contribution to the energy budget has been identified in imaging and spectroscopic observations 
and interpreted in terms of Alfv\'enic wave energy (\citealp{TOMetal2007}, \citealp{MCIetal2011}, \citealp{THUetal2014}). 
Alfv\'enic waves have long been assumed to play a significant role in plasma heating, since 
their incompressible nature enables 
them to transfer energy over large distances. Their potentially significant role in determining the nature of the solar
wind has long been known, with observations of Alfv\'enic fluctuations observed from early in-situ measurements 
(\citealp{BELDAV1971}) and the Alfv\'enicity of the fluctuations evident over many frequency decades (periods 
from seconds to days - e.g., \citealp{BRUCAR2005}). The \textit{Coronal Multi-channel Polarimeter} (CoMP) was the first 
instrument to provide evidence for the Alfv\'enic
wave energy flux through the solar atmosphere (\citealp{TOMetal2007}), and subsequent investigation has revealed the
ubiquity and persistence of this wave flux in Doppler velocities (\citealp{TOMMCI2009}; \citealp{DEMetal2014}; 
\citealp{MORetal2015, MORetal2016}). However because CoMP is ground-based, the afforded observing window 
only allows for measurements of three frequency decades (seconds to an hour, Fig~\ref{fig:features}).

These observations have come in conjunction with some success in producing a heating of coronal plasma from wave-driven 
models (e.g., \citealp{SUZINU2005}; \citealp{CRAetal2007}; \citealp{EVAetal2009}) along with 
reproducing some of the basic properties of slow and fast solar wind. However, there are a number of challenges that these 
models have to overcome (\citealp{CRA2009}, \citealp{OFM2010}), requiring more stringent observational constraints on the 
mechanisms for delivering energy, mass and momentum (EMM) into the source regions in the low corona. Moreover, 
current forecasting 
models employ empirical techniques that have provided relative levels of success (\citealp{LEEetal2009}), 
although they are limited in their predictive power due 
to the neglect of the physical mechanisms ultimately responsible for plasma heating and wind acceleration. For example, 
the Wang, Sheeley \& Arge (WSA) model relies upon a static, potential coronal field and an empirical formula to 
estimate wind 
speed, while the Magnetohydrodynamics-Around-a-Sphere (MAS) model generates a wind by adding an ad-hoc heating function. 
Successful advancement of our knowledge of the underlying physics and improving forecasting abilities will depend 
on the accuracy of
determining the EMM fluxes through different magnetic regions, quantifying the relative contributions of the plethora of 
potential mechanisms and detailed knowledge of the free energy in the coronal magnetic field.

Initial results from CoMP have shown promise for synoptic imaging-spectroscopy of the extended corona to contribute to our 
understanding of the nature and evolution of the coronal magnetic field and associated dynamics. Such an instrument has significant advantages 
over slit spectrometers (e.g., \textit{Hinode} EIS) that can only provide focused observations of portions of the corona, 
and also over synoptic imagers (\textit{Solar Dynamic Observatory} (SDO) Atmospheric Imaging Assembly (AIA)) that only provide 
broadband intensity measurements.
 In particular, recent CoMP results demonstrate the capability
to make unique insights into global behaviour of Alfv\'enic wave phenomena through the solar corona (Figure~\ref{fig:features}), with the potential to 
constrain key features of models, e.g., wave excitation \& damping/mode conversion; Alfvenic turbulence; 
relative energy fluxes through distinct regions of the corona; energy flux from the lower solar atmosphere to the solar 
wind (\citealp{TOMMCI2009}; \citealp{VERTHetal2010}; \citealp{DEMetal2014}; \citealp{MORetal2015, MORetal2016}). 
Additionally, CoMP has demonstrated 
imaging-spectrometer coronagraphs have the potential for the exploitation of waves through magneto-seismology, in combination with 
spectroscopic techniques, to determine local plasma conditions, e.g., measurements of the plane-of-sky component of the
magnetic field and propagation angle 
with respect to the solar surface. Furthermore, this combination can also provide estimates of the outflow of plasma low in the 
corona (\citealp{MORetal2015}), which potentially allows for constraints to be placed on EMM fluxes 
along open-field lines and could also contribute to the identification of regions contributing to the slow solar wind.

CoMP has demonstrated the potential for imaging-spectrometer coronagraphs to reveal unique insights into Alfv\'enic 
waves and the ability to constrain their contribution to coronal heating and wind acceleration, but there are some things 
unachievable by ground-based observations. For example, ground-based instruments will be unable to provide the necessary 
extended sequences of observations required to probe the Alfv\'enic waves over the broad frequency range observed out in 
the solar wind. Additionally,
ground-based instruments will also be unable to provide the coverage necessary for continual forecasting of Space 
Weather due to seeing, weather and the day-night cycle.  
\smallskip

\section{Science Goals}

The following provides only a small selection of potential science and operational questions that could be answered by a 
space-based imaging-spectrometer coronagraph. They are mainly focused on the study of wave phenomenon and energy
transfer via Alfv\'enic waves. However, a coronagraph would also be able to contribute to many further science questions 
surrounding dynamics and ejecta in the corona, e.g., Coronal Mass Ejections (\citealp{TIAetal2013}).

\begin{enumerate} 
\item	What are the key physical mechanisms contributing to coronal heating in different magnetic geometries?\\
(i)	What is the relative Alfv\'enic wave energy flux through different magnetic regions?\\
(ii)	What are the physical rates for energy deposition by Alfv\'enic waves in the solar corona? \\
(iii)	Is there evidence for the development of Alfv\'enic wave turbulence in the lower corona?
\item	What is the role of Alfv\'enic waves in the acceleration of the solar wind?\\
(i)	What is the evolution of these waves between the solar corona and the solar wind?\\
(ii)	How does the Alfv\'enic wave energy flux vary over the course of the solar cycle?
\item	Which regions are key contributors to solar wind streams?
\item	Is it feasible to exploit MHD waves via magneto-seismology to provide routine and meaningful characterisation 
of the plasma and magnetic field conditions in the corona?
\end{enumerate} 

\smallskip

\section{Requirements \& Justification}

\noindent\textit{Observables:} In order to make progress in answering the above science questions, a number of diagnostics 
are required. The envisaged capabilities and observables of a space-based imaging-spectrometer draw directly from the heritage
of the CoMP instrument, although technology from other imaging spectrometers will also be valuable. CoMP has been 
making routine measurements of the 1074.7~{nm} (Fe XIII) line, taking images at three 
wavelength positions (1074.50~{nm}, 1074.62~{nm}, 1074.74~{nm}) with a 0.13 nm FWHM filter bandpass. Imaging data products 
are provided with a 30~s cadence and a 
spatial sampling of 4.46" over a 1.05-1.3~$R_{\sun}$ field of view. While the spatial, temporal and spectral sampling are 
relatively coarse compared to other ground- and space-based instruments, the data products still reveal a wealth of coronal 
dynamics in the line profile diagnostics, e.g.,
line core intensity, Doppler shifts, line widths (e.g., \url{http://mlso.hao.ucar.edu/mlso_data_COMP_2016.php} ). This,
coupled with the large synoptic FOV, has enabled unique insights into the global evolution of the coronal magnetic field 
and coronal Alfv\'enic waves. 

The suitability of the Iron XIII infra-red emission lines for probing coronal physics has been
discussed in a number of papers, e.g., \cite{JUD1999} and demonstrated by CoMP. The infra-red Iron line is spectrally
broad enough to allow sampling across the line profile, as opposed to narrower EUV coronal lines. In addition, 
the ability to observe the 1074.7~{nm} and 1079.8~{nm} Fe XIII line pair has proved 
particularly beneficial. The line pair is density sensitive and has enabled co-spatial and co-temporal estimates
of the coronal electron density, a key diagnostic for understanding EMM fluxes.  

Ideally, increasing the spatial, temporal and spectral resolution of the instrument, compared to CoMP, would prove 
beneficial. An instrument with similar throughput and sensitivity to CoMP would benefit from the lower noise and
scattering afforded by being space-based, permitting improvements in each of these areas. The benefit of 
higher cadences ($5-10$~s) and spatial sampling ($\sim$1") would allow a much finer sampling of the coronal dynamics, 
extending the range of phenomenon that can be studied, e.g., study of the higher-frequency Alfv\'enic waves glimpsed in 
both SDO/AIA and the High-resolution Coronal Imager ($\sim$60~s - \citealp{MORMCL2013}; \citealp{THUetal2014}). This
would also avoid the under-resolution of the spectroscopic diagnostics from line profiles currently found in the CoMP 
observations (affecting Doppler velocities and line widths, e.g. \citealp{MCIDEP2012}), enabling increased accuracy and
sensitivity for wave measurements (demonstrated in EIS wave studies - \citealp{VANetal2008c}). Additionally, finer spectral sampling would allow for an improvement in the 
uncertainties of spectral line profiles. And finally, a larger FOV out to
2~$R_{\sun}$ would also provide a significant increase in capability, enabling the evolution of dynamics and ejecta
through the low corona to be studied.\\

\noindent\textit{Why observations must be done from space?:} Current data sets from CoMP suffer from various 
restrictions due to being taken from the ground, and there would be significant benefits from placing a similar instrument in space. Current ground-based observations
are subject to a narrow time window before seeing conditions become an issue, 
with only few data-sets achieving lengths of 3 hours, which hampers the study of events over long time-scales 
(e.g., Alfv\'enic waves over extended frequency ranges). Major constraints also exist on when data can be taken and 
the quality of taken data due to weather conditions and pollutants in the atmosphere. These factors then restrict the
capability to provide daily data products required for Space Weather monitoring and forecasting, and also creates 
problems for long term studies of the evolution of the EMM fluxes over the solar cycle. 

In addition, Earth's atmosphere also creates significant problems, with variable seeing conditions and increased scattering of photons leading
to lower signal to noise ratios. Furthermore, atmospheric absorption lines also pose problems for accurate photospheric
continuum measurements, which are a critical part of the data preparation. As such, these  issues have meant that
CoMP has not been able to provide routine measurements of large coronal holes or the extended corona due to
poor signal to noise, despite having the potential to do so.

Perhaps more importantly, there will soon be a need for a space-based coronagraph. The ageing SOHO/Lasco and 
STEREO/Cor 2 are 
still being relied upon and there is currently no planned mission with a coronagraph that could provide continued
monitoring of the corona. Such an instrument is relied upon, not only by the science community, but also by the 
international forecasting agencies who exploit the intensity images to provide constraints on the size and direction 
of CMEs. A low corona (1-2~$R_{\sun}$) imaging-spectrometer in combination with a wide FOV (\textgreater2~$R_{\sun}$), 
white light imager would provide a unique combination to study the initiation and development of dynamics in the low 
corona and their impact on, and evolution out into, the heliosphere (e.g., \citealp{DEFetal2014,DEFetal2016}).

\bibliographystyle{aa}
\bibliography{mybib}

\begin{thebibliography}{23}
\expandafter\ifx\csname natexlab\endcsname\relax\def\natexlab#1{#1}\fi

\bibitem[{{Belcher} \& {Davis}(1971)}]{BELDAV1971}
{Belcher}, J.~W. \& {Davis}, Jr., L. 1971, \jgr, 76, 3534

\bibitem[{{Bruno} \& {Carbone}(2005)}]{BRUCAR2005}
{Bruno}, R. \& {Carbone}, V. 2005, Living Reviews in Solar Physics, 2, 4

\bibitem[{{Cranmer}(2009)}]{CRA2009}
{Cranmer}, S.~R. 2009, Living Reviews in Solar Physics, 6, 3

\bibitem[{{Cranmer} {et~al.}(2007){Cranmer}, {van Ballegooijen}, \&
  {Edgar}}]{CRAetal2007}
{Cranmer}, S.~R., {van Ballegooijen}, A.~A., \& {Edgar}, R.~J. 2007, \apjs,
  171, 520

\bibitem[{{De Moortel} {et~al.}(2014){De Moortel}, {McIntosh}, {Threlfall},
  {Bethge}, \& {Liu}}]{DEMetal2014}
{De Moortel}, I., {McIntosh}, S.~W., {Threlfall}, J., {Bethge}, C., \& {Liu},
  J. 2014, \apj, 782, L34

\bibitem[{{DeForest} {et~al.}(2014){DeForest}, {Howard}, \&
  {McComas}}]{DEFetal2014}
{DeForest}, C.~E., {Howard}, T.~A., \& {McComas}, D.~J. 2014, \apj, 787, 124

\bibitem[{{DeForest} {et~al.}(2016){DeForest}, {Matthaeus}, {Viall}, \&
  {Cranmer}}]{DEFetal2016}
{DeForest}, C.~E., {Matthaeus}, W.~H., {Viall}, N.~M., \& {Cranmer}, S.~R.
  2016, \apj, 828, 66

\bibitem[{{Evans} {et~al.}(2009){Evans}, {Opher}, {Jatenco-Pereira}, \&
  {Gombosi}}]{EVAetal2009}
{Evans}, R.~M., {Opher}, M., {Jatenco-Pereira}, V., \& {Gombosi}, T.~I. 2009,
  \apj, 703, 170

\bibitem[{{Judge}(1998)}]{JUD1999}
{Judge}, P.~G. 1998, \apj, 500, 1009

\bibitem[{{Lee} {et~al.}(2009){Lee}, {Luhmann}, {Odstrcil}, {MacNeice}, {de
  Pater}, {Riley}, \& {Arge}}]{LEEetal2009}
{Lee}, C.~O., {Luhmann}, J.~G., {Odstrcil}, D., {et~al.} 2009, \solphys, 254,
  155

\bibitem[{{McIntosh} \& {De Pontieu}(2012)}]{MCIDEP2012}
{McIntosh}, S.~W. \& {De Pontieu}, B. 2012, \apj, 761, 138

\bibitem[{{McIntosh} {et~al.}(2011){McIntosh}, {de Pontieu}, {Carlsson},
  {Hansteen}, {Boerner}, \& {Goossens}}]{MCIetal2011}
{McIntosh}, S.~W., {de Pontieu}, B., {Carlsson}, M., {et~al.} 2011, \nat, 475,
  477

\bibitem[{{Morton} \& {McLaughlin}(2013)}]{MORMCL2013}
{Morton}, R.~J. \& {McLaughlin}, J.~A. 2013, \aap, 553, 10

\bibitem[{{Morton} {et~al.}(2015){Morton}, {Tomczyk}, \& {Pinto}}]{MORetal2015}
{Morton}, R.~J., {Tomczyk}, S., \& {Pinto}, R. 2015, Nature Communications, 6,
  7813

\bibitem[{{Morton} {et~al.}(2016){Morton}, {Tomczyk}, \& {Pinto}}]{MORetal2016}
{Morton}, R.~J., {Tomczyk}, S., \& {Pinto}, R. 2016, \apj, 828, 89

\bibitem[{{Ofman}(2010)}]{OFM2010}
{Ofman}, L. 2010, Living Reviews in Solar Physics, 7, 4

\bibitem[{{Suzuki} \& {Inutsuka}(2005)}]{SUZINU2005}
{Suzuki}, T.~K. \& {Inutsuka}, S.-i. 2005, \apj, 632, L49

\bibitem[{{Thurgood} {et~al.}(2014){Thurgood}, {Morton}, \&
  {McLaughlin}}]{THUetal2014}
{Thurgood}, J.~O., {Morton}, R.~J., \& {McLaughlin}, J.~A. 2014, \apj, 790, L2

\bibitem[{{Tian} {et~al.}(2013){Tian}, {Tomczyk}, {McIntosh}, {Bethge}, {de
  Toma}, \& {Gibson}}]{TIAetal2013}
{Tian}, H., {Tomczyk}, S., {McIntosh}, S.~W., {et~al.} 2013, \solphys, 288, 637

\bibitem[{{Tomczyk} \& {McIntosh}(2009)}]{TOMMCI2009}
{Tomczyk}, S. \& {McIntosh}, S.~W. 2009, \apj, 697, 1384

\bibitem[{{Tomczyk} {et~al.}(2007){Tomczyk}, {McIntosh}, {Keil}, {Judge},
  {Schad}, {Seeley}, \& {Edmondson}}]{TOMetal2007}
{Tomczyk}, S., {McIntosh}, S.~W., {Keil}, S.~L., {et~al.} 2007, Science, 317,
  1192

\bibitem[{{van Doorsselaere} {et~al.}(2008){van Doorsselaere}, {Nakariakov},
  {Young}, \& {Verwichte}}]{VANetal2008c}
{van Doorsselaere}, T., {Nakariakov}, V.~M., {Young}, P.~R., \& {Verwichte}, E.
  2008, \aap, 487, L17

\bibitem[{{Verth} {et~al.}(2010){Verth}, {Terradas}, \&
  {Goossens}}]{VERTHetal2010}
{Verth}, G., {Terradas}, J., \& {Goossens}, M. 2010, \apj, 718, L102

\end{thebibliography}

\end{document}